\documentclass[english]{article}
\usepackage[T1]{fontenc}
\usepackage[latin9]{inputenc}
\usepackage{amsmath}
\usepackage{amssymb}
\usepackage{stmaryrd}
\usepackage{graphicx}
\usepackage{setspace}
\usepackage{esint}



\usepackage{babel}
\begin{document}
\begin{onehalfspace}
\begin{center}
{\large{}Set of Holonomic and Protected Gates on Topological Qubits
for Realistic Quantum Computer}
\par\end{center}{\large \par}
\end{onehalfspace}

\begin{center}
{\footnotesize{}Andrey R. Klots$^{1}$, Lev B. Ioffe$^{1,2}$}
\par\end{center}{\footnotesize \par}

\begin{doublespace}
{\footnotesize{}$^{1}$Department of Physics, University of Wisconsin
-- Madison, Madison, WI 53706 USA}{\footnotesize \par}

{\footnotesize{}$^{2}$Google Inc., Venice, CA 90291 USA}{\footnotesize \par}
\end{doublespace}

\noindent \textbf{\footnotesize{}Keywords: }{\footnotesize{}Holonomic
Gate, Geometric Gate, Protected Gate, Protected Qubit, High-fidelity
Gate, Superconducting Qubit}{\footnotesize \par}

\textbf{Abstract}. In recent years qubit designs such as transmons
approached the fidelities of up to 0.999. However, even these devices
are still insufficient for realizing quantum error correction requiring
better than 0.9999 fidelity. Topologically protected superconducting
qubits are arguably most prospective for building a realistic quantum
computer as they are intrinsically protected from noise and leakage
errors that occur in transmons. We propose a topologically protected
qubit design based on a $\pi$-periodic Josephson element and a universal
set of gates: protected Clifford group and highly robust (with infidelity
$\sim10^{-4}$) non-discrete holonomic phase gate. The qubit is controlled
via charge($Q$) and flux($\Phi$)-biases. The holonomic gate is realized
by quickly, but adiabatically, going along a particular closed path
in the two-dimensional $\{\Phi,Q\}$-space \textendash{} a path where
computational states are always degenerate, but Berry curvature is
localized inside the path. This gate is robust against currently achievable
noise levels. This qubit architecture allows building a realistic
scalable superconducting quantum computer with leakage and noise-induced
errors as low as $10^{-4}$, which allows performing realistic error
correction codes with currently available fabrication techniques. 

\begin{figure}
\includegraphics[width=0.8\columnwidth]{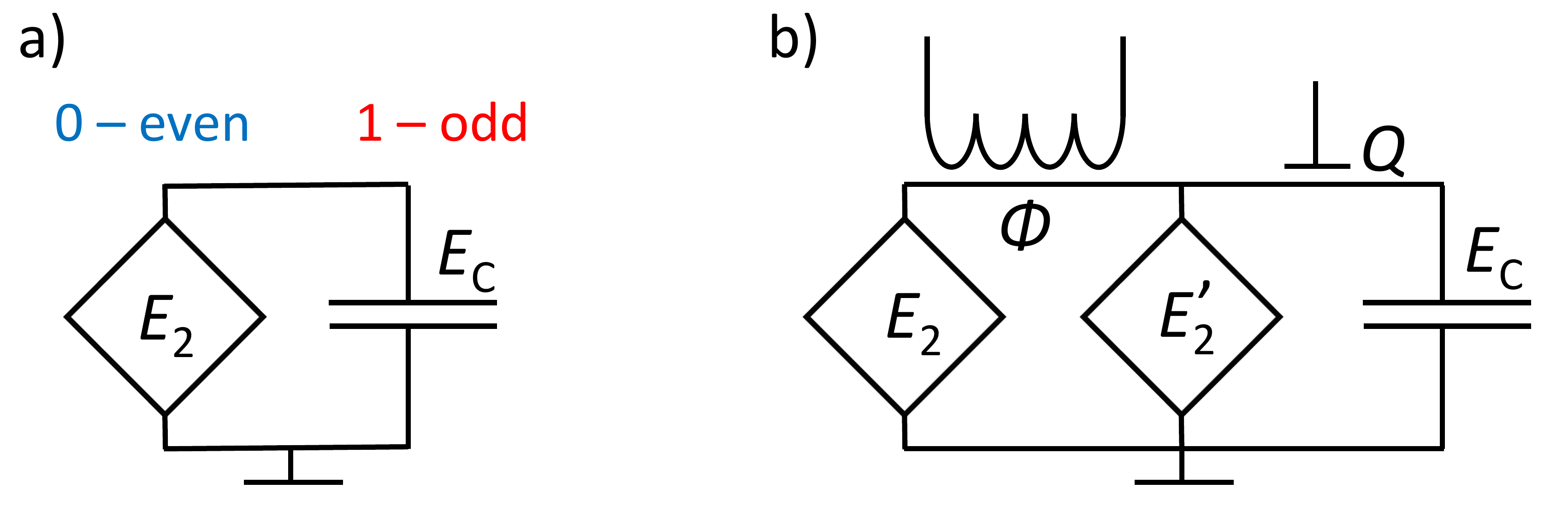}

\protect\caption{\textbf{Schematics of the protected $\pi-$periodic qubit} (a) and
its modification that allows full set of operations (b). \label{fig:Schematics} }
\end{figure}

\textbf{Overview}. The main challenge in building a realistic quantum
computer is building a qubit and developing logical operations that
can be used for efficient error corrections\cite{SURFCODEversluis2017scalable}.
The problem with currently existing qubits is two fold. First, the
best fidelity of the existing qubits is either insufficient for error
correction or requires an impractically large hardware overhead. Second,
the transmons displaying the best fidelity achieve it by reducing
non-linearity, increasing leakage out of computational space that
is very difficult to correct within surface code\cite{LEAKAGEaliferis2005fault}.
With increasing complexity of the quantum algorithm, surface codes
require rapidly increasing number of physical qubits to perform error
correction. As an alternative, it is desirable to create qubits that
are protected against noise on the hardware level. Arguably the most
prospective design involves using a $\pi$-periodic element\cite{RHOMBbell2014protected,RHOMBgladchenko2009superconducting,RHOMBioffe2002possible,RHOMBgroszkowski2018coherence,SIGATEbrooks2013protected}
\textendash{} effectively a Josephson element that only allows tunneling
of even number of Cooper pairs and has phase($\varphi$)-energy($E$)
relation $E=-E_{2}\cos2\varphi$ with $E_{2}$ being the Josephson
energy for double Cooper pair tunneling. Such an element coupled to
a large capacitor $C$ with charging energy $E_{C}\ll E_{2}$ (here
$E_{C}=(2e)^{2}/2C$) forms the qubit (Fig. \ref{fig:Schematics}a)
in which two logical states are characterized by the charge parity
(i.e. parity of number of Cooper pairs) on the superconducting island:
``0'' and ``1'' logical states are encoded by even and odd charge
states respectively. The dephasing rate for such qubit is exponentially
suppressed with increasing value of $\sqrt{E_{2}/E_{C}}$, which is
a square of the characteristic width of the wavefunction $\psi(n)$
in the charge ($n$) space. In this respect, the $\pi-$periodic qubit
is similar to the transmon where protection is achieved due to the
exponential suppression of the energy dispersion as a function of
the charge offset ($Q$). However, unlike transmons, $\pi-$periodic
qubits are strongly anharmonic and have nearly degenerate computational
states well separated from excited ones, preventing leakage outside
of the computational space.

Ideally, a protected qubit should allow a universal set of fault tolerant
operations during which the qubit remains protected. Here the term
``protected'' implies exponential suppression of any noise and term
``robust'' \textendash{} suppression of linear noise. In this paper
we show that a relatively minor modification of the $\pi-$periodic
qubit gives an almost ideal protected qubit. Namely, it allows fault-tolerant
(i.e. with exponentially small error) $Z(\frac{\pi}{2})$ discrete
phase gate and robust non-discrete holonomic phase gate $Z(\Theta)$
along with previously proposed\cite{SIGATEbrooks2013protected} $X(\frac{\pi}{2})$-
and $X\varotimes X(\frac{\pi}{2})$-gates. Altogether these gates
allow universal qubit control\cite{CLIFFORDbravyi2005universal}.
During all these operations the qubit states remain degenerate. Whilst
this degeneracy is exponentially protected during discrete gates,
it is only insensitive in the linear order to the charge noise for
holonomic phase gate. Furthermore, due to degeneracy of the computational
states the holonomic operation is not sensitive to a precise form
of the pulse shape in time domain. Because holonomic operations are
robust but not exponentially protected the resulting qubit is \emph{almost
ideal. }

The modification that gives almost ideal qubit is the ability to vary
the value of the Josephson energy, $E_{2}$ together with the offset
charge, $Q$. Thus, we control two parameters, and a closed path in
the 2D parameter space produces Berry phase of the qubit. Variation
of the effective $E_{2}$ can be achieved by replacing a $\pi$-periodic
element by a dc-SQUID-like loop of two similar $\pi$-periodic elements
connected in parallel (Fig. \ref{fig:Schematics}b). We refer to this
circuit as $\pi$-SQUID with effective Josephson energy $E_{2}^{\textrm{eff}}$
depending on flux $\Phi$ through the loop. For $\Phi=0$, effective
$E_{2}^{\textrm{eff}}$ is the largest and the qubit behaves like
a regular protected $0-\pi$ qubit. Increasing $\Phi$ decreases $E_{2}^{\textrm{eff}}$.
When $\sqrt{E_{2}^{\textrm{eff}}/E_{C}}\lesssim1$ a relatively slow
variation of the parameters (the estimate will be given below) squeezes
the qubit wavefunction to only one or two charge states and protection
is lifted. This temporary removal of protection creates a strong charge($Q$)-dispersion
and allows to perform different phase gates.

\begin{figure}
\includegraphics[width=1\columnwidth]{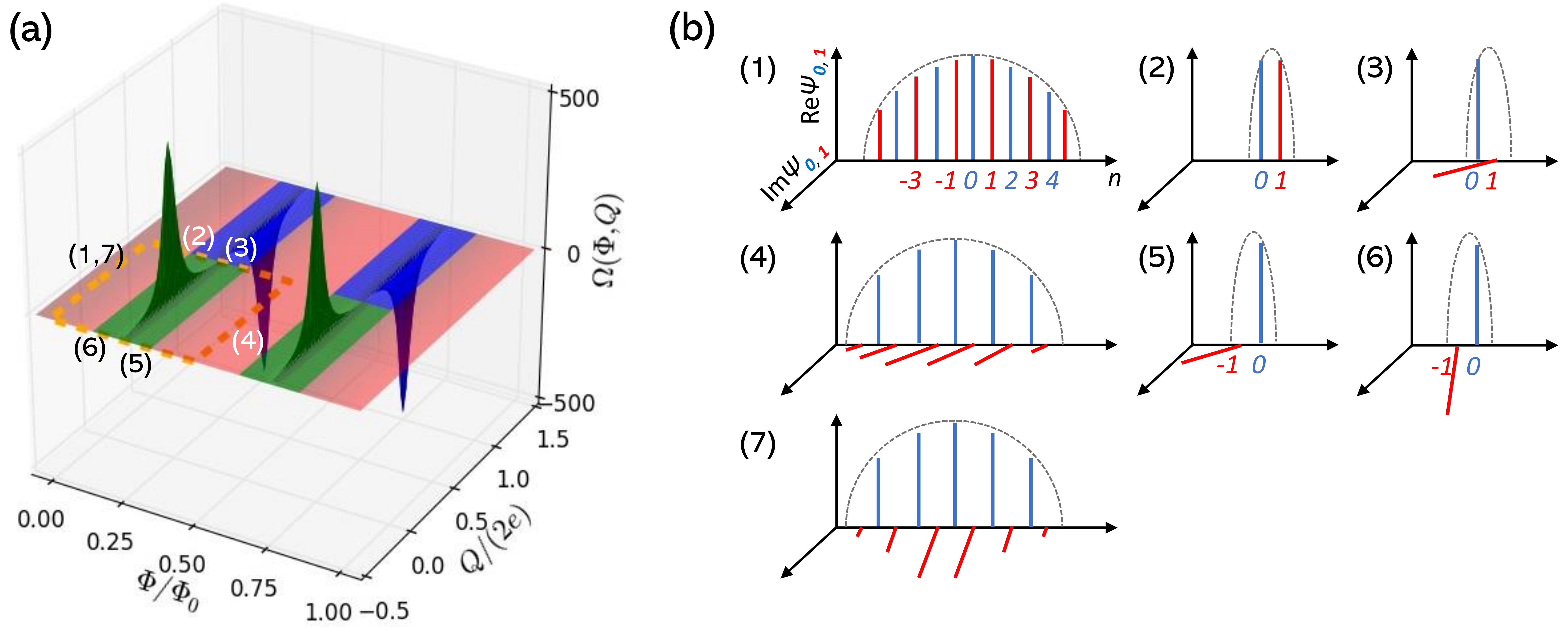}

\protect\caption{\textbf{Berry curvature and phase.} a) Berry curvature in parameter
space $\{\Phi,Q\}$. The protected region is shown by red color, unprotected
by green and blue. The holonomic transformation is achieved by changing
the parameters along dashed orange line that starts at the point (1)
in the protected regime. b) Cartoons of the wave function at a few
characteristic points in parameter space. Points 1, 4, 7 correspond
to protected regime, points 2, 3, 5, 6 to unprotected one, see text.
\label{fig:Berry}}
\end{figure}

We show below that flux and charge bias variables $\{\Phi,Q\}$ form
a 2D parameter space, in which the qubit possesses a Berry curvature
shown in Fig. \ref{fig:Berry}a that is obtained analytically. The
Berry curvature has a strong peak at $(\Phi=\Phi_{0}/4,Q=0)$. The
non-discrete phase gate is performed by adiabatically going in a loop
around this peak and gaining different Berry phases for the two logical
states. It is crucial that one can chose the path so that at every
point the computational states remain degenerate and Berry curvature
is zero. The preservation of degeneracy implies that the gate is holonomic:
lifting of protection does not cause a gain of unwanted time-dependent
dynamic phase. Furthermore, it provides exponential protection against
flux noise, while the symmetric nature of the $Q=1/2$ point implies
the absence of the linear response to charge noise. Throughout the
paper we use units of $\hbar=2e=1$ and $\Phi_{0}=2\pi$.

The discrete protected $Z(\frac{\pi}{2})$ phase gate is performed
by turning the $\pi$-SQUID off thereby allowing the qubit to evolve
only under the quadratic capacitor Hamiltonian $H_{C}=E_{C}n^{2}$.
Analogously to the $X(\frac{\pi}{2})$-gate\cite{SIGATEbrooks2013protected},
by choosing the proper gate timing we can make even states to to gain
0 dynamic phase and all odd states \textendash{} dynamic phase of
$-\pi/2$. Similarly to the $X(\frac{\pi}{2})$-gate this transformation
is protected and the errors are flagged by the qubit excitation to
a high energy state. 

\textbf{Qualitative description of the holonomic gate}. Figure \ref{fig:Berry}a
shows the path (dashed orange loop) along which the holonomic gate
is performed. Red areas of the figure depict protected regions where
$\Phi$ is close to $0$ (or $\pi$) and $\sqrt{E_{2}^{\textrm{eff}}(\Phi)/E_{C}}\gg1$.
Outside that region ($\sqrt{E_{2}^{\textrm{eff}}/E_{C}}\lesssim1$)
the qubit is in an unprotected regime.

The loop can be split into two branches: top ($Q>0$) and bottom ($Q<0$).
First we go along the $Q>0$-branch applying a positive charge bias
on the superconducting island while keeping the qubit protected (Fig.
\ref{fig:Berry}a, (1)). Then, increasing the flux through the loop,
we lift the protection against dephasing and squeeze the even wavefunction
into only one charge state of $n=0$ and odd wavefunction \textendash{}
to $n=1$ (2,3). Importantly, the flux through the $\pi$-SQUID loop
also creates a gauge transformation that rotates each charge state
$n$ on the island by a different phase factor $n\tilde{\Phi}$, where
$\tilde{\Phi}$ is some gauge-related rotation angle that will be
discussed and derived further. As a result, on the $Q>0$-branch the
$n=0$-state remains unaffected, but the strongest odd state (Fig.
\ref{fig:Berry}b. (2-3)) $n=+1$ gains the phase of $(+1)\tilde{\Phi}$.
Then we return to our initial state through the $Q<0$-branch (Fig.
\ref{fig:Berry}b. (4-7)). On this branch the gauge transform is performed
in the opposite direction ($\tilde{\Phi}\rightarrow-\tilde{\Phi}$).
The dominant even state $n=0$ is again unaffected, but the odd state,
now represented by the $n=-1$ charge state, gains the phase of $(-1)(-\tilde{\Phi})$
which has the same sign as on the $Q>0$-branch. Thus, on both halves
of the path the odd state is rotated in the same direction causing
a non-discrete rotation that is smaller than $\pi$ by a value proportional
to the asymmetry of the $\pi$-SQUID.

\textbf{Quantitative description of the holonomic gate}. First, let
us discuss the properties of the Hamiltonian and its eigenfunctions.
We focus on the designs of $\pi$-periodic elements where computational
energy levels are well separated from excited ones\cite{RHOMBbell2014protected,RHOMBgladchenko2009superconducting}.
Energy of the $\pi$-squid is $E(\varphi)=-E_{2}\cos(\varphi-\Phi/2)-E_{2}'\cos(\varphi+\Phi/2)$.
In this case the relevant low energy degrees of freedom of the qubit
are described by the Hamiltonian:

\begin{equation}
\mathcal{H}=E_{C}\left(n-Q\right)^{2}-E_{2}^{eff}\left(\Phi\right)\cos2\left(\varphi-\tilde{\Phi}\left(\Phi\right)\right).\label{eq:Hshort}
\end{equation}

Here

\begin{equation}
E_{2}^{\textrm{eff}}\left(\Phi\right)=\sqrt{E_{2}^{2}+E_{2}'^{2}+2E_{2}E_{2}'\cos2\Phi};\label{eq:EofPhi}
\end{equation}

\begin{equation}
\tilde{\Phi}\left(\Phi\right)=\frac{1}{2}\arctan\left(\frac{\sin\Phi}{E_{2}+E_{2}'},\frac{\cos\Phi}{E_{2}'-E_{2}}\right),\label{eq:PsiofPhi}
\end{equation}
with Josephson energies $E_{2}$ and $E_{2}'$ for the two $\pi$-periodic
elements. We choose parameters so that $E_{2}+E_{2}'\gg E_{C}\gg|E_{2}-E_{2}'|$.
This allows to have both protected ($E_{2}^{\textrm{eff}}(0)/E_{C}\gg1$)
and unprotected ($E_{2}^{\textrm{eff}}(\pi/2)/E_{C}\ll1$) regimes.
The eigenfunctions $\psi(n)$ of such Hamiltonian are represented
by either even or odd charge states enclosed by the envelope function
of width $\sim(E_{2}^{\textrm{eff}}/E_{C})^{1/4}$ that is centered
around $n=Q$ in the charge space (Fig.2b.1-2). Note that for half-integer
$Q$ even and odd eigenstates have a mirror symmetry, which means
that for half-integer $Q$ computational states are degenerate regardless
of $\Phi$. Thus, offset charge $Q$ changes the balance between even
and odd states. Flux bias, in turn, changes the width of the wavefunction
by modifying $E_{2}^{\textrm{eff}}$. However, flux bias has another
crucial effect: although it is tempting to disregard the phase offset
$\tilde{\Phi}$ in (\ref{eq:Hshort}), it should not be done because
in our gate $\tilde{\Phi}$ is a function of $\Phi$, which is not
constant. In fact, this term plays a key role in realizing the gate:
since the potential in (\ref{eq:Hshort}) is shifted by $\tilde{\Phi}$
in the $\varphi$-space, the wavefunction $\psi(\varphi)$ is also
transformed as $\psi(\varphi)\rightarrow\psi(\varphi-\tilde{\Phi})$.
In the charge space this results in a gauge transformation $\psi(n)\rightarrow\exp(in\tilde{\Phi})\psi(n)$,
mentioned in the previous section.

\textbf{Calculating Berry curvature.} The phase accrued by the computational
states when going along the loop is calculated as an integral of the
Berry curvature over the area enclosed by the loop. In this section
we derive the expression for the Berry curvature.

We begin with the region of $\Phi\approx\pi/2$ where $E_{2}^{\textrm{eff}}\lesssim E_{C}$
and the computational eigenfunctions are squeezed to only one or two
charge states (Fig. \ref{fig:Berry}, (2)). For any value of $Q$
it is convenient to write the effective Hamiltonian in terms of the
two charge states $n$ nearest to $Q$. For example, for $0<Q<1$
the odd state can be written in the basis of two wavefunctions $\psi_{\pm1}(\varphi)=(2\pi)^{-1/2}\exp(\pm i\varphi)$,
corresponding to $n=\pm1$ charge states, while the relevant even
states are $\psi_{0}(\varphi)=(2\pi)^{-1/2}$ and $\psi_{2}(\varphi)=(2\pi)^{-1/2}\exp(2i\varphi)$:

\begin{align}
\mathcal{H}_{\text{eff}}^{\text{odd}} & \approx-\frac{1}{2}E_{2}^{\text{eff}}\left(\sigma^{x}\cos2\tilde{\Phi}+\sigma^{y}\sin2\tilde{\Phi}\right)+QE_{C}\sigma^{z},\label{eq:Hodd}\\
\mathcal{H}_{\text{eff}}^{\text{even}} & \approx-\frac{1}{2}E_{2}^{\text{eff}}\left(\sigma^{x}\cos2\tilde{\Phi}+\sigma^{y}\sin2\tilde{\Phi}\right)+(1-Q)E_{C}\sigma^{z}.\label{eq:Heven}
\end{align}
Here $\sigma^{x,y,z}$ are Pauli matrices in the space of $\left|+1\right\rangle $,
$\left|-1\right\rangle $ charge state vectors in (\ref{eq:Hodd})
, and in the basis of $\left|0\right\rangle $, $\left|2\right\rangle $
in (\ref{eq:Heven}). Ground states of $\mathcal{H}_{\text{eff}}^{\text{even}}$
and $\mathcal{H}_{\text{eff}}^{\text{odd}}$ represent even and odd
computational states respectively. Higher eigenstates of $\mathcal{H}_{\text{eff}}^{\textrm{odd}(\textrm{even})}$
lie outside of the computational space. For the path shown in Fig.
\ref{fig:Berry}a the excited states always remain separated from
the computational states by a large energy gap, so that effect of
these excited states can be ignored. For $-\frac{1}{2}\lesssim Q\lesssim\frac{1}{2}$
the accumulated phase is only due to odd eigenstates. To avoid the
excitations of the higher energy states of $\mathcal{H}_{\textrm{eff}}^{\textrm{odd}(\textrm{even})}$
the gate speed needs to be much slower than the smallest energy gap
between the eigenvalues of $\mathcal{H}_{\text{eff}}^{\textrm{odd}(\textrm{even})}$
on the path, i.e. 
\begin{equation}
\tau_{\text{gate}}^{-1}\ll E_{C}\label{eq:1/tau_gate}
\end{equation}
 where $\tau_{\text{gate}}$ is time of the gate operation. Analytically,
probability of Landau-Zener tunneling from even(odd) computational
state to the even(odd) excited state can be evaluated from the Hamiltonian
(\ref{eq:Hodd},\ref{eq:Heven}) as\cite{LZshytov2003landau,LZzener1932non}
\begin{equation}
P=\exp[-\tau_{\text{gate}}E_{C}^{2}/(E_{2}+E_{2}')].\label{eq:ProbLZ}
\end{equation}

The ground state of (\ref{eq:Hodd}) is described by a spinor $\left|\textrm{spinor}\right\rangle =(\begin{array}{cc}
e^{-i\xi/2}\cos\theta/2, & e^{+i\xi/2}\sin\theta/2\end{array})^{T}$. Polar and azimuthal angles of the spinor are related to the qubit
parameters as 
\begin{equation}
\begin{array}{l}
\xi=2\tilde{\Phi}(\Phi);\\
\theta=\textrm{\ensuremath{\arctan}}\left(-\frac{E_{2}^{eff}(\Phi)}{2},E_{C}Q\right).
\end{array}\label{eq:SpinorMap}
\end{equation}

\begin{figure}
\includegraphics[width=1\columnwidth]{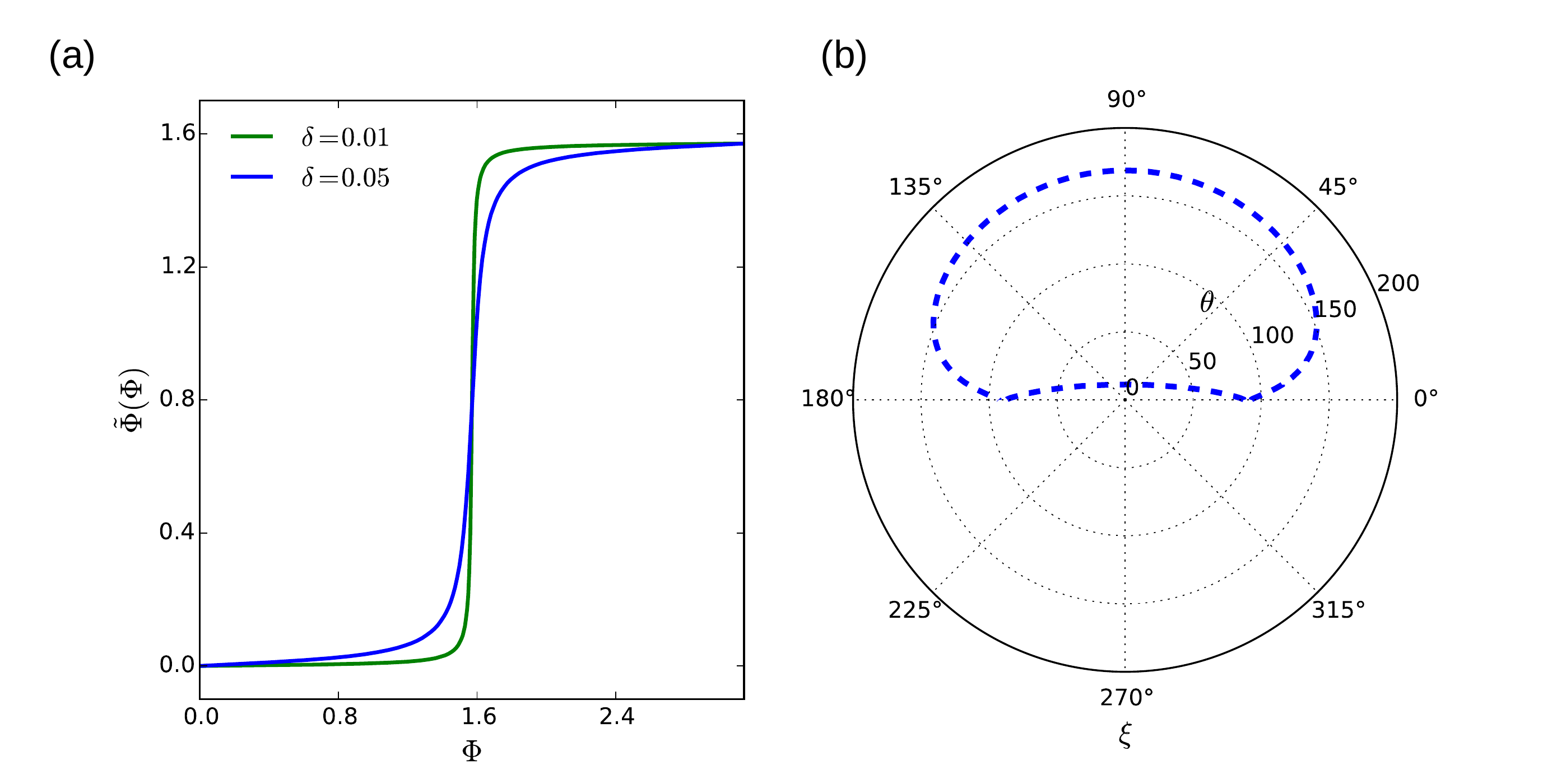}

\protect\caption{\textbf{Coordinate transformations.} a) Mapping of $\Phi$ onto $\tilde{\Phi}$
for $\delta=0.01,0.05$. b) Coordinate transformation that relates
the loop in the qubit parameter space $\Phi$ and $Q$ to the effective
field acting on the spin, characterized by Euler angles $\xi$ and
$\theta$: mapping of our loop onto the Bloch sphere. \label{fig:Mapping}}
\end{figure}

In order to obtain the Berry curvature we perform a coordinate transform
($\Phi=\Phi(\xi),Q=Q(\theta,\xi)$) and map the well-known Berry curvature
of a spin $\Omega_{\xi\theta}^{\textrm{spin}}=(1/2)\sin\theta$ onto
variables $(\Phi,Q)$ as $\Omega_{\Phi Q}^{\textrm{eff}}=\frac{\partial\xi}{\partial\Phi}\frac{\partial\theta}{\partial Q}\Omega_{\xi\theta}^{\textrm{spin}}$.
Since $\frac{\partial\tilde{\Phi}}{\partial\Phi}$ and $\frac{\partial\xi}{\partial\Phi}$
has a maximum at $\Phi=\pi/2$ (Fig.3a), most of the Berry curvature
is concentrated in the vicinity of half-integer values of $\Phi/\pi$.
This allows us write equations in the vicinity of $\Phi=\Phi_{0}/4=\pi/2$.
Let us introduce new notations $\alpha\overset{\textrm{def}}{=}\Phi-\pi/2$,
$E_{\Sigma}\overset{\textrm{def}}{=}E_{2}+E_{2}'$, $\delta\overset{\textrm{def}}{=}|E_{2}'-E_{2}|/E_{\Sigma}$
and $e_{C}\overset{\textrm{def}}{=}2E_{C}/E_{\Sigma}$. For large
$\alpha\gg e_{C}$ equations (\ref{eq:Hodd}, \ref{eq:Heven}) break
down but this regime gives little contribution to the Berry phase
because in it the Berry curvature, along with the qubit charge dispersion
is exponentially suppressed with $(\alpha/e_{C})^{1/2}$. 

The dimensionless parameter that controls the Berry phase accumulated
in the adiabatic evolution is 
\[
\eta=\delta/e_{C}=|E_{2}'-E_{2}|/2E_{C}.
\]
In the following we assume that $\eta\ll1$. The protection is removed
when $\alpha\lesssim e_{C}\ll1$, and restored when $\alpha\gg e_{C}$.
In the former regime the adiabatic evolution leads to accumulation
of significant Berry phase. From (\ref{eq:EofPhi}) we approximate
$E_{2}^{\text{eff}}(\alpha)\approx E_{\Sigma}\sqrt{\delta^{2}+\alpha^{2}}$
and Berry curvature for the Hamiltonian (\ref{eq:Hodd}) reduces to
a simple form

\begin{equation}
\Omega_{\Phi Q}^{\text{peak}}\left(\frac{\pi}{2}+\alpha,Q\right)=\frac{e_{C}}{2}\frac{\delta}{\left(\delta^{2}+\alpha^{2}+\left(e_{C}Q\right)^{2}\right)^{3/2}}\left\{ 1+\mathcal{O}\left(\alpha^{2}\right)\right\} .\label{eq:OmegaEff}
\end{equation}

The total Berry curvature (difference between $\Omega_{\Phi Q}^{\text{peak}}$
for even and odd states) that determines the phase difference gained
between odd and even states can be evaluated as

\begin{equation}
\Omega_{\Phi Q}\left(\Phi,Q\right)=\left\{ \begin{array}{ll}
\Omega_{\Phi Q}^{\text{peak}}\left(\Phi,Q\right)-\Omega_{\Phi Q}^{\text{peak}}\left(\Phi,Q-1\right) & ,Q>0\\
\Omega_{\Phi Q}^{\text{peak}}\left(\Phi,Q\right)-\text{\ensuremath{\Omega}}_{\Phi Q}^{\mbox{peak}}\left(\Phi,Q+1\right) & ,Q<0
\end{array}\right..\label{eq:OmegaTot}
\end{equation}
Since $\Omega_{\Phi Q}$ is an odd function of $Q-1/2$ it is equal
to zero at half-integer values of $Q$. This expression holds for
$|Q|\le1/2$ and $0\le\Phi\le\pi$, but can be generalized to the
entire $\{\Phi,Q\}$-space by keeping in mind that $\Omega_{\Phi Q}$
has a period of $\pi$ in $\Phi$ and period of $2$ in $Q$. Also,
at half-integer values of $Q$ the computational states are degenerate.
Additionally, as mentioned above, the Berry curvature and energy splitting
between the two lowest states is exponentially suppressed for $\Phi\approx0$
and $\Phi\approx\pi$. Thus, we choose our holonomic adiabatic path
to go through these regions of $\Phi=0,\pi$ and $Q=\pm1/2$.

\textbf{Berry phase}. Let us evaluate the the Berry phase that is
given by integral of the Berry curvature (\ref{eq:OmegaEff}) over
$\alpha$ and $Q$ in the leading approximation in $\eta\ll1$. The
integral is dominated by the region $\{|\alpha|\lesssim\delta,|e_{C}Q|\lesssim\delta\}$.
For $\delta\ll e_{C}$ this implies that $\alpha\ll e_{C},Q\ll1$
which justifies the use of (\ref{eq:OmegaEff}). If we were to integrate
the curvature (\ref{eq:OmegaEff}) in the infinite limits of $\alpha$
and $Q$ we would get $\Theta_{0}=\pi$ for the Berry phase. However,
exponential suppression of curvature for $\alpha>e_{C}$ and finite
size of the loop limited by $Q\approx\pm1/2$ implies that actual
Berry phase is given by the integral that is cut off in these directions:

\begin{equation}
\Theta\approx\intop_{\sim-e_{C}}^{\sim e_{C}}d\alpha\intop_{\approx-e_{C}/2}^{\approx e_{C}/2}d(e_{C}Q)\frac{\delta/2}{\left(\delta^{2}+\alpha^{2}+\left(e_{C}Q\right)^{2}\right)^{3/2}}\approx\pi-\mathcal{A}\frac{\delta}{e_{C}},\label{eq:Theta}
\end{equation}
giving us leading approximation in $\eta=\delta/e_{C}$. This simple
analytical computation does not give the value of the constant $\mathcal{A}\sim1$
which we determined numerically. Our numerical calculations done by
diagonalizing the Hamiltonian (\ref{eq:Hshort}) in the basis of 201
charge-state wavevector $\left\{ \left|-100\right\rangle ,\left|-99\right\rangle ,...,\left|+100\right\rangle \right\} $
and going along the contour with step of $0.001\pi$ in $\Phi$ and
$0.01$ in $Q$ determine the numerical constant $\mathcal{A}=2.97\pm0.02$.

Most importantly, the resulting phase is not discrete: it deviates
from a discrete rotation of $\pi$ by a value that can be controlled
by tuning the qubit design. For a reasonably achievable values of
$\delta\sim10^{-2}$ and $e_{C}\sim10^{-1}$, the non-discrete rotation
is $\mathcal{A}\eta\sim0.3$rad. Non-discreteness of this gate can
be understood by mapping the rectangular path $\gamma$ in the $\{\Phi,Q\}$-space
onto a Bloch sphere $\{\xi,\theta\}$ using equations (\ref{eq:SpinorMap}).
As shown in figure 3b, the path covers an area, which is somewhat
less than half of the Bloch sphere. Moreover, our numerical modeling
shows that further increasing of $\eta$ can yield any non-discrete
phase (Fig.4a) from $\Theta\approx\pi$ for $\eta\ll1$ to $\Theta\rightarrow0$
for $\eta\gtrsim1$ \textendash{} regime when qubit does not leave
the protected state ($E_{2}^{\mbox{eff}}(\forall\Phi)\gg E_{C}$)
and $\Omega_{\Phi Q}$ remains exponentially small. 

\begin{figure}
\includegraphics[width=1\columnwidth]{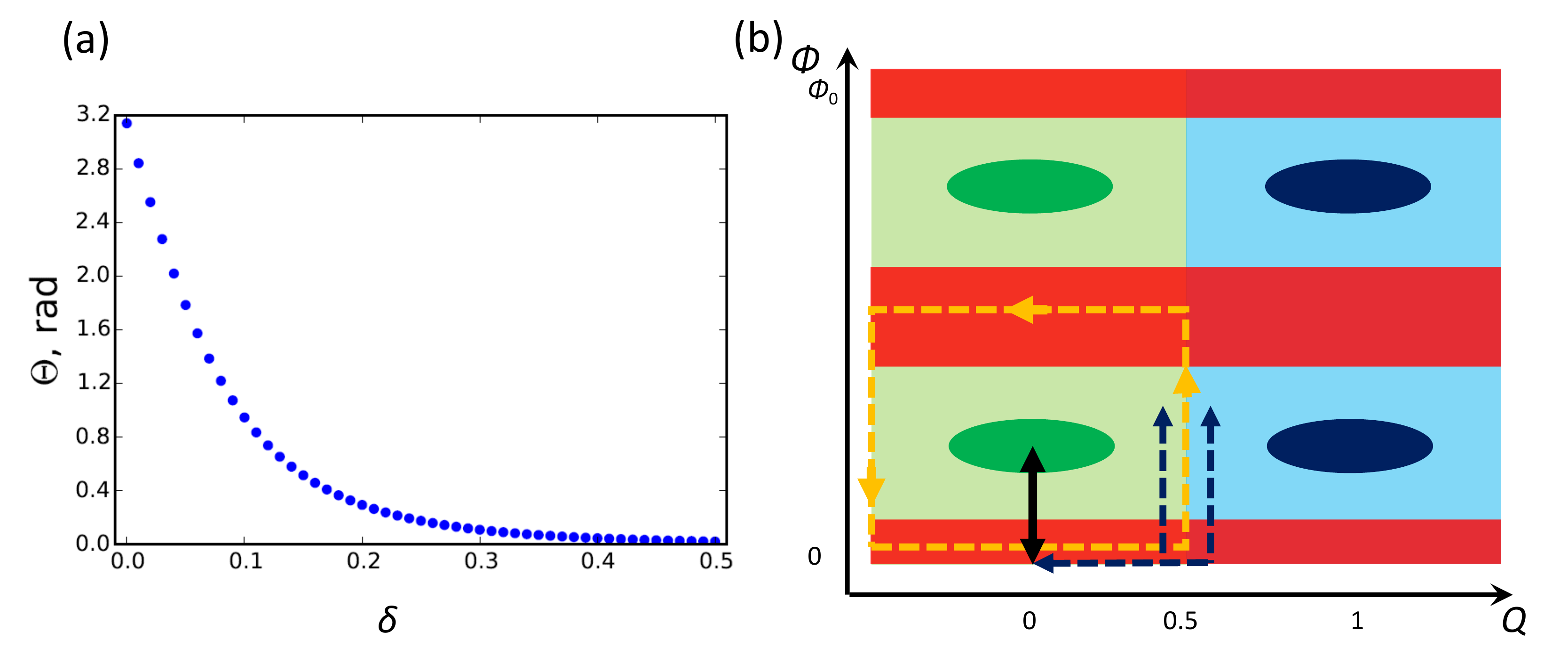}

\protect\caption{\textbf{Phase gates.} a) Phase $\Theta$ as a function of the $\pi$-SQUID
asymmetry $\delta$ assuming $e_{C}=0.1$. b) Sketch of the Berry
curvature map with depiction of paths that realize different gates.
Orange dashed line: rectangular adiabatic loop to realize the holonomic
phase gate. Black solid line: quick diabatic $X(\pi/2)$-gate realized
by quick frustration of the $\pi$-SQUID. Dark-blue dashed line: example
of an idle gate that can be used in CPMG or multiple echo sequences.
During the idle gate only accrual of the positive or negative noise-induced
dynamic phase occurs. This can be used to create a multiple charge-echo
effect to partially compensate the random dynamic phase accrued during
the holonomic gate. \label{fig:Realization}}
\end{figure}

\textbf{The relative gate error and timing.} Since the part of the
path sensitive to the flux noise ($\Phi\approx0,\pi$; $-\frac{1}{2}\le Q\le\frac{1}{2}$)
is located in the protected regime of $E_{2}^{\textrm{eff}}/E_{C}\gg1$,
effects of the flux noise are exponentially suppressed. In order to
estimate the effect of the charge noise, assume that horizontal (i.e.
in $\Phi$-direction) part of the path is shifted vertically by a
small value $\epsilon_{Q}$, so that $Q=1/2+\varepsilon_{Q}$ instead
of $Q=1/2.$ This path would give Berry phase that differs from (\ref{eq:Theta})
by 
\begin{equation}
\begin{array}{ll}
\Delta\Theta & \sim\int_{\sim-e_{C}}^{\sim+e_{C}}d\alpha\int d\epsilon_{Q}\Omega_{\Phi Q}\left(\frac{\pi}{2}+\alpha,\frac{1}{2}+\epsilon_{Q}\right)\sim\\
 & \sim\int\eta\epsilon_{Q}d\epsilon_{Q}\sim\eta\epsilon_{Q}^{2}.
\end{array}\label{eq:ChargeError}
\end{equation}

Here we used the fact that $\Omega_{\Phi Q}\sim\eta\epsilon_{Q}$
is linear with $\epsilon_{Q}$ near $Q=\pm1/2$. Thus, from (\ref{eq:Theta})
relative error of the nondiscrete part ($\Theta\mod\pi$) of the phase
rotation $\Theta$ is $\varepsilon_{\Theta}^{\textrm{rel}}\sim\Delta\Theta/\eta\sim\varepsilon_{Q}^{2}$.
Assuming a high value of the charge noise $\varepsilon_{Q}\sim10^{-2}$
on few-second timescales\cite{CHARGEchristensen2019anomalous,CHARGEastafiev2004quantum,CHARGEastafiev2006temperature,CHARGESEThenning1999bias},
we arrive at a relative gate error as low as $\varepsilon_{\Theta}^{\textrm{rel}}\sim10^{-4}$. 

In order for the gate to be considered adiabatic to a reasonable degree,
we want to have the previously estimated probability of Landau-Zener
tunneling (\ref{eq:ProbLZ}) out of the computational space to be
$P=\exp[-\tau_{\textrm{gate}}e_{C}^{2}E_{\Sigma}/4]\sim10^{-4}$.
For reasonable values $e_{C}\sim10^{-1}$ and $E_{\Sigma}\sim2\pi\times40$GHz\cite{RHOMBbell2014protected},
we get the gate timing $\tau_{\textrm{gate}}\gtrsim15$ns.

Finally, we consider an error due to accumulation of unwanted dynamic
phase. We consider this to be the bottleneck problem for any protected
qubit design because in order to perform a non-discrete rotation one
needs to temporarily remove the protection by either (a) lifting the
degeneracy of the computational states, which leads to error linear
with error in gate timing or (b) keep the degeneracy of the computational
states (like in our case) at a cost of gaining linear dispersion of
the computational states which leads to error linear in noise amplitude.
In our gate, for example, the qubit remains in unprotected regime
$\{\Phi\approx\frac{\pi}{2}\pm e_{C},Q=\pm\frac{1}{2}\}$ during the
time $\tau_{\textrm{u}}\sim e_{C}\tau_{\textrm{gate}}$, when it gains
dynamic phase $\gamma$. There the computational states are degenerate
but their charge dispersion $\Delta E_{1,0}(\epsilon_{Q})$ is linear
with deviation $\epsilon_{Q}$ of charge offset from $Q=\pm0.5$.
At $\Phi=\pi/2$ the dispersion is $\Delta E_{1,0}(\epsilon_{Q})=\textrm{sign}(Q)E_{\Sigma}e_{C}\epsilon_{Q}$.
Notably, for $Q>0$ and $Q<0$ parts of the path the dynamic phase
has opposite sign. Such accumulation of dynamic phase is identical
to charge echo experiments\cite{CHARGEECHOnakamura2002charge,CHARGEastafiev2004quantum,NOISEMATHnesterov2012modeling}
which are sensitive only to high-frequency ($\mbox{f}\gtrsim\tau_{\textrm{u}}^{-1}$)
noise. Approximating the ``turn-on'' function for the unprotected
regime as a square pulse we can characterize the dynamic phase by
its mean square using a well-known expression\cite{CHARGEECHOnakamura2002charge,NOISEMATHnesterov2012modeling}:

\[
\overline{\gamma^{2}}\sim\int d\omega\Delta E_{1,0}^{2}\frac{A}{\omega}\left(\frac{\sin(\omega\tau_{\textrm{u}}/2)}{\omega/2}\right)^{2}\sim A\left(\Delta E_{10}\tau_{\textrm{u}}\right)^{2}.
\]

Here $A/\omega$ is the spectral density of 1/f-noise. We can now
estimate the infidelity of the phase gate. Assume that the ideal gate
acting on initial qubit state $\left|\textrm{initial}\right\rangle $
gives the state $\left|\textrm{ideal}\right\rangle =Z(\Theta)\left|\textrm{initial}\right\rangle $.
The physical gate gives instead a state $\left|\textrm{real}\right\rangle =Z(\Theta+\gamma)\left|\textrm{initial}\right\rangle $
with $\gamma\ll1$. Define the mean infidelity as $1-F=1-\overline{\left|\left\langle \textrm{ideal}\left|\textrm{real}\right.\right\rangle \right|}\approx\overline{\gamma^{2}}/2$.
Assuming same parameters as above, $E_{\Sigma}\sim2\pi\times40$GHz;
$e_{C}=0.1$, $\tau_{\textrm{gate}}=15$ns and high-frequency 1/f
charge noise with amplitude $\epsilon_{Q}=A^{1/2}$ between $1.5\times10^{-4}\times(2e)$
and $6.5\times10^{-4}\times(2e)$\cite{CHARGEECHOnakamura2002charge,CHARGEastafiev2004quantum,CHARGEQUANTDOTfujisawa2000charge,CHARGEELECTROMETERzimmerli1992noise}
we get $1-F$ between $\sim10^{-4}$ and $\sim10^{-5}$. This estimate
relies on the assumption that charge noise follows 1/f dependence
up to $\text{GHz frequencies, similar to the }$charge sensitive devices
studied in works by Astafiev et al.\cite{CHARGEastafiev2004quantum}\textbf{
}and expected theoretically by Faoro et al.\cite{CHARGETHEORfaoro2006quantum}.
Notice that the dephasing can be decreased even further by controlling
the qubit using GRAPE\cite{GRAPEchasseur2015engineering} pulses to
further minimize gate timing and by using CPMG-like\cite{CPMGbylander2011noise}
or multiple-echo sequences to filter out 1/f noise (Fig.4b). This
would give us further improvement in fidelity. To the best of our
knowledge, this gate performance is much better than in any currently
existing qubit\cite{FIDELITYgong2019genuine,FIDELITYneill2018blueprint,FIDELITYsheldon2016characterizing},
especially considering a complexity of non-discrete gates.

\textbf{Discrete fault-tolerant $Z(\pi/2)$-gate.} The proposed qubit
can also be used to perform a protected discrete $\exp(-i\pi\sigma^{z}/4)$-gate.
The idea is similar to the gate proposed in work by Brooks et al.\cite{SIGATEbrooks2013protected}.
In contrast to the nondiscrete gate, this gate is not performed adiabatically,
but by quick modification of the Hamiltonian. While the adiabatic
change of $\Phi$ leads to squeezing of the wavefunction in the charge
space, in the case of the quick modification of the Hamiltonian, the
wavefunction does not have time to squeeze, but it remains as it was,
delocalized in the charge space. In more detail, we start with the
qubit in the protected state with $\Phi=0$ and $E_{2}^{\mbox{eff}}=E_{\Sigma}$.
Then we quickly change the flux to $\Phi=\pi/2$. This effectively
turns off the $\pi$-SQUID leaving only the capacitive part of the
Hamiltonian $\mathcal{H}(\pi/2)=n^{2}/2C+\mathcal{O}(\delta E_{\Sigma}\cos2\varphi)$
and hence the qubit evolves under the operator $U_{\pi/2}(t)\approx\exp\{-in^{2}(2C)^{-1}t\}$.
After time $T=\pi C$ qubit is brought back into the initial state.
With this gate timing $U_{\pi/2}(T)\approx\exp\{-i\pi n^{2}/2\}$.
As a result, all even charge states are multiplied by a factor of
$\exp\left\{ -i\pi n_{\mbox{even}}^{2}/2\right\} =1$ and odd states
-- by $\exp\left\{ -i\pi n_{\mbox{odd}}^{2}/2\right\} =-i$. Hence,
we realize a $\exp(-i\pi\sigma^{z}/4)$-gate. This gate is dual to
the gate proposed in the work by Brooks el al.\cite{SIGATEbrooks2013protected}
and therefore has similar exponential stability against gate timing
error $T\rightarrow\pi C+\Delta T$ (see section VI of \cite{SIGATEbrooks2013protected})
and perturbation stability (e.g. against small perturbation $\mathcal{O}(\delta E_{\Sigma}\cos2\varphi)$
in the Hamiltonian: see section XI of work by Brooks et al.\cite{SIGATEbrooks2013protected}).
In both cases errors and perturbations only result into excitations
of high-energy levels that are outside of our computational space.

Since in a protected state ($\Phi=0$) our qubit is identical to a
standard $0-\pi$ qubit\cite{RHOMBgroszkowski2018coherence,SIGATEbrooks2013protected}
it is also possible to implement a $\exp(i\pi\sigma^{x}/4)$-gate
described by Brooks et al.\cite{SIGATEbrooks2013protected}. With
these two discrete gates it is possible to realize the topologically
protected Clifford group $\mathcal{C}_{1}$ and a two-qubit gate\cite{SIGATEbrooks2013protected}
$X\varotimes X(\frac{\pi}{2})$. In combination with the semiprotected
holonomic gate described above it results in a universal qubit control\cite{CLIFFORDbravyi2005universal}
with high fidelity.

\textbf{Conclusion.} We showed that by adding one more degree of freedom
to the protected qubit architecture based on double periodic Josephson
junctions it is possible to realize two more types of gates: a discrete
protected gate and a robust continuous holonomic gate that is not
sensitive to the flux noise and to charge noise in the linear order.
Together with previously existing one- and two-qubit flip-gate\cite{SIGATEbrooks2013protected}
it is possible to build a first realistic scalable quantum computer
with universal qubit control and infidelity of the order of $10^{-4}/$gate
(importantly, with potential for further improvement). For the holonomic
gate, in principle, one can also choose a different, more complicated
path in the $\{\Phi,Q\}$-space and achieve different phase gates
with different noise sensitivity. This diversity arises due to non-trivial
Berry curvature landscape of an essentially two-dimensional system
that is controlled by two bias channels. We expect that by creating
more complex circuits with more degrees of freedom one can create
systems with more complex Berry curvature landscapes and gauge fields.
In prospective it will be interesting to generalize this approach
to other types of protected or robust qubits such as fluxonium\cite{FLUXONIUMCOHERENCEnguyen2018high,FLUXONIUMmanucharyan2009fluxonium,FLUXONIUMT1lin2018demonstration}.
We hope that further development of similar holonomic qubit architectures
will allow achieving higher degrees of protection for continuous gates. 

\textbf{Acknowledgements}. We thank Robert F. McDermott, Lara Faoro
and Bradley Christensen from University of Wisconsin -- Madison for
useful discussions and comments. This work is supported by the U.S.
Government under Grant No. W911NF-18-1-0106.


\end{document}